      [1999/12/01 v1.4c Il Nuovo Cimento]
\documentclass{cimento}

\usepackage{graphicx}  
\title{NICMOS Measurements of the Near Infrared Background}
\author{R.~Thompson,
D.~Eisenstein,
X.~Fan,
M.~Rieke\from{ins:a}\\
        \atque
R.~Kennicutt\from{ins:c}}
\instlist{\inst{ins:a} Steward Observatory, University of Arizona, Tucson AZ, USA,
  \inst{ins:c} Cambridge University, Cambridge UK}

\PACSes{\PACSit{95.75.De}
\PACSit{---.---}{\ldots}}

\begin{document}

\maketitle

\begin{abstract}
This paper addresses the nature of the near infrared background.  We investigate
whether there is an excess background at 1.4 microns, what is the source of the 
near infrared background and whether that background after the subtraction of all 
known sources contains the signature of high redshift objects ($Z > 10$).  Based 
on NICMOS observations in the Hubble Ultra Deep Field and the Northern Hubble Deep 
Field we find that there is no excess in the background at 1.4 microns and that the
claimed excess is due to inaccurate models of the zodiacal background. We find that
the near infrared background is now spatially resolved and is dominated by galaxies 
in the redshift range between 0.5 and 1.5. We find no signature than can be 
attributed to high redshift sources after subtraction of
all known sources either in the residual background or in the fluctuations of
the residual background.  We show that the color of the fluctuations from both
NICMOS and \emph{Spitzer} observations are consistent with low redshift objects and
inconsistent with objects at redshifts greater than 10. It is most likely that
the residual fluctuation power after source subtraction is due to the outer 
regions of low redshift galaxies that are below the source detection limit and
therefore not removed during the source subtraction.
\end{abstract}

\section{Introduction}

The nature of the near infrared background is the subject of intense current
investigation.  Much of this interest centers on whether the near infrared 
background contains the signature of very high redshift ($Z > 10$) sources.
Claims for such a signature have been spurred by the claim of an excess in
the background with a sharp cutoff to the blue of 1.4$\mu$ \cite{ref:mat05} 
and from fluctuation analyses of deep \emph{Spitzer} data \cite{ref:kas05,ref:kas07b}. 
These findings are in contrast to earlier analyses of NICMOS images from the 
Northern Hubble Deep Field (NHDF) \cite{ref:thm03} and the Hubble Ultra Deep 
Field (HUDF) \cite{ref:thm06} where no excess was found.

\section{The Near Infrared Background Excess}

Observations with Infrared Telescope in Space \cite{ref:mat05} found a near 
infrared background at 1.4$\mu$m of 70 nw m$^{-2}$
sr$^{-1}$ after subtraction of modeled zodiacal flux and modeled contributions
from stars and galaxies. Observations in the NHDF and hUDF \cite{ref:mad00,
ref:thm03,ref:thm06} measured a total contribution
from stars and galaxies of 7-12 nw m$^{-2}$ sr$^{-1}$ after
subtraction of a zodiacal background measured from a median of all of the images
taken in the field. A later analysis \cite{ref:thm07a} showed that the total 
powers measured in both investigations were essentially identical, therefore the
discrepancy was not due to instrumental effects.  The difference is in the 
subtracted zodiacal light.  The measured zodiacal light in the NICMOS images
is greater than the modeled zodiacal light in \cite {ref:mat05} by almost 
exactly the claimed excess.  This led to the conclusion that the claimed
excess did not exist and is the result of the inadequacy of the zodiacal
models to accurately predict the zodiacal flux.  The error in the model was
relatively modest ($28\%$), however, since the source flux is so small,
$2\%$ of the zodiacal flux, the error led to a significant excess of flux
not due to zodiacal or known sources.  The analysis in \cite{ref:thm07a}
removes the false 1.4$\mu$ excess as evidence for a high redshift component
to the observed near infrared background.

\section{Nature of the Near Infrared Background}

The sources in the zodiacal subtracted NICMOS images in the NHDF and NUDF 
provide all of the measured power.  Photometric redshifts \cite{ref:thm03,
ref:thm06} show that the majority of power is provided by galaxies in the
redshift range between 0.5 and 2.0.  From these measurements we conclude
that the observed near infrared background is now resolved into galaxies
and is primarily due to galaxies at relatively low redshifts.  This 
extragalactic background is a small percentage of the overall background
which is due to zodiacal reflected emission from nearby dust.

\section{The Source Subtracted Background}

Having determined that the near infrared background is due to resolved galaxies
in the redshift range between 0.5 and 2.0 we can then ask about the nature of
the background after all known resolved sources have been subtracted.  The 
source positions and extents were determined by an optimal extraction technique
\cite{ref:sza99} that utilizes both the ACS and NICMOS images in all of the 
six bands.  The source subtracted image was then produced by setting all of 
the pixels identified as being a source to zero.  This procedure only removed
$7\%$ of the pixels from the image.  Note that the method in \cite{ref:sza99}
is a single pixel criterion.  In the extraction we utilized SExtractor 
\cite{ref:ber96} to impose the additional criterion that a source must contain
at least 3 contiguous pixels.  Neither of these techniques extends the source
beyond the region where a single pixel meets the source detection criterion.
This is important in the analysis that follows.

\subsection{Fluctuation Analysis}

\begin{figure}
\includegraphics[totalheight=0.5\textheight]{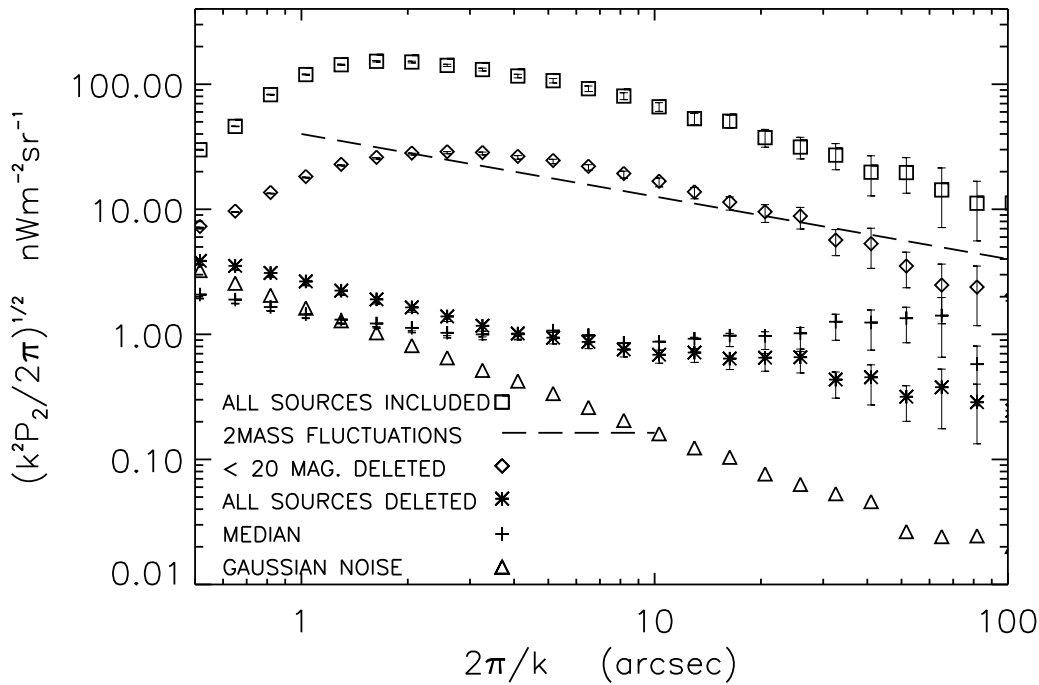}     
\caption{The fluctuation spectrum of the of the F160W NUDF image is given
by the squares, the image with sources brighter than 20 AB mag. subtracted by
the diamonds, with all sources subtracted by the asterisks, and the fluctuations
of a Gaussian noise field by the plus signs. The dashed line represents an
average of the fluctuations found by \cite{ref:kas02} in 7 different 2MASS
calibration fields.  The photon Poisson noise for the all sources included
and brighter than 20. mag deleted curves is smaller than the symbol sizes.
The noise in the all sources deleted curve can be estimated from the deviations
from a smooth curve. \label{fig-fluct}}
\end{figure}

We use a fluctuation analysis to investigate whether the residual background
after source subtraction contains a signal from a distribution of sources that
are fainter than the single source detection limit.  The fluctuation analysis
is based on a 2 dimensional Fourier transform of the background image and is 
described in detail in Apendix A of \cite{ref:thm07a}.  The results of
the analysis on the F160W HUDF image are shown in Figure~\ref{fig-fluct}.  The
figure for the F110W images is essentially identical.  It is clear from the difference
at longer wavelengths between the all sources subtracted power and the gaussian
noise power that there is signal in the residual background after source 
subtraction. 

\section{Nature of the Fluctuation Sources} 

Several studies have attributed the residual fluctuations after source subtraction
to very high redshift ($z>10$) sources.  Fluctuations in deep 2MASS calibration
images have been attributed to high redshift sources \cite{ref:kas02} as well as 
fluctuations in deep \emph{Spitzer} images \cite{ref:kas05,ref:kas07b}. We will 
address the 2MASS and \emph{Spitzer} images separately.

\subsection{2MASS Fluctuations}

In \cite{ref:kas05} all detectable sources in 7 deep H band 2MASS calibration field 
images were subtracted out down to an equivalent AB magnitude of 20 and a fluctuation
analysis performed on each of the images.  The average of the fluctuations in the
images is shown as the dashed line in Figure~\ref{fig-fluct}.  To test whether the
remaining fluctuations were due to high redshift objects we subtracted sources in
the NICMOS F160W HUDF images down to the same limiting magnitude.  Only 10 sources out of
the 5000 sources in the NICMOS image were at or brighter than the subtraction limit
in the 2MASS images.  The fluctuations from the 20th magnitude or brighter subtracted 
image are shown by the diamonds in Figure~\ref{fig-fluct}.  They are consistent with
the 2MASS fluctuations over the common region of spatial wavelengths.  Next all of 
the sources were subtracted in the NICMOS image and the analysis performed again.
The result is shown by the asterisks.  The all source subtracted fluctuations
are significantly below the 20th magnitude or brighter subtracted fluctuations 
indicating that the observed sources in the much deeper NICMOS image can easily
account for the fluctuations.  All of these sources have redshifts less than 7
and the predominant power comes from sources in the redshift range between 0.5 and
2.0 as would be expected from the analysis of the sources that provide the 
majority of the near infrared background power.  The conclusion is that the 
observed fluctuations in the 2MASS source subtracted image are due to low redshift
objects below the 2MASS detection limit and do not indicate the presence of
very high redshift objects.  Details of this analysis are given in \cite{ref:thm07a}.

\section{HUDF and \emph{Spitzer} Fluctuations}

We next turn to the fluctuations in the HUDF \cite{ref:thm07a,ref:thm07b} and
\emph{Spitzer} \cite{ref:kas05,ref:kas07a,ref:kas07b} images to see if they
require the presence of high redshift sources.

\subsection{HUDF Fluctuations}

\begin{figure}
\includegraphics[totalheight=0.5\textheight]{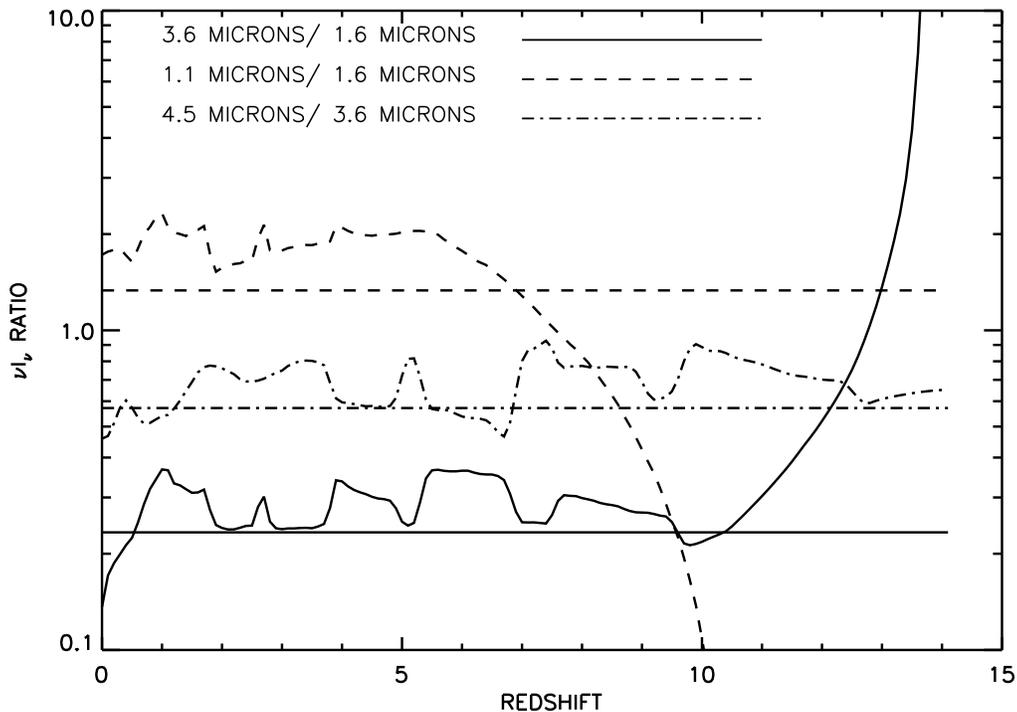}     
\caption{The expected ratios of fluctuations in the NICMOS and \emph{Spitzer}
bands are shown in the figure versus redshift. The \emph{Spitzer} 3.6 micron
to NICMOS F160W ratio is given by the solid line, the NICMOS F110W to NICMOS
F160W ratio by the dashed line and the \emph{Spitzer} 4.5 micron to \emph{Spitzer}
3.6 micron ratio by the dash dot line.  In each case the flat horizontal line
gives the observed value. \label{fig-ratio}}
\end{figure}

In addition to the fluctuations in the F160W NICMOS HUDF image shown in
Figure~\ref{fig-fluct} we also analyzed the F110W image for fluctuations.  The
spatial spectrum of the F110W fluctuations are almost identical to the F160W
fluctuations.  Using the predominant SEDs in the HUDF we next calculated the
expected ratio of fluctuations in the F110W and F160W NICMOS bands versus
redshift as well as the \emph{Spitzer} 3.6 and 4.5 micron bands.  The results
of the calculation are shown in Figure~\ref{fig-ratio}.  The observed ratio of
the NICMOS band fluctuations given by the horizontal dashed line is inconsistent
with sources with redshifts greater than eight so we conclude that the fluctuations
in the source subtracted NICMOS images are not due to high redshift sources.
We consider the most likely source of the residual fluctuations to be the 
extended regions of the observed sources that were too faint to be detected by our 
source subtraction technique.

\subsection{Spitzer Fluctuations}

We show in \cite{ref:thm07b} that the degree of source subtraction in the NICMOS
HUDF and the \emph{Spitzer} images used in \cite{ref:kas07a,ref:kas07b} are 
essentially equal and it is therefore legitimate to make a comparison.  Due to
the long wavelength of the \emph{Spitzer} bands the Lyman break does not enter
them even for redshifts as high as 15 so their ratio is not a sensitive indicator
of the redshift of the residual fluctuations.  The ratio of the NICMOS F160W
to \emph{Spitzer} 3.6 micron fluctuations, however, indicates that the ratio is
incompatible with redshifts above 10.  We therefore conclude that the residual
fluctuations in the \emph{Spitzer} images are not evidence for the presence of
very high redshift objects.

\subsection{Comparison of Spatial Wavelength Spectra}

\begin{figure}
\includegraphics[totalheight=0.5\textheight]{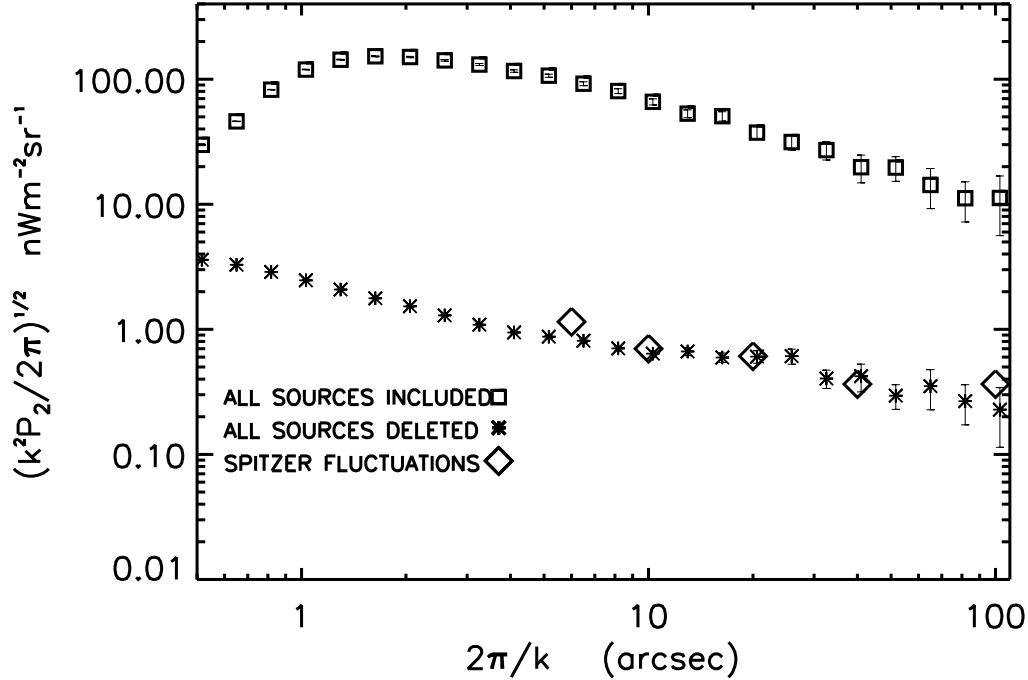}     
\caption{A comparison between the observed residual NICMOS F160W fluctuations,
asterisks and the \emph{Spitzer} 3.6 micron fluctuations at spatial scales
of 5 arc seconds and greater.  Within the noise they are identical. 
\label{fig-space}}
\end{figure}

The point was made in \cite{ref:kas07b} that the spatial spectrum of the residual
fluctuations in the \emph{Spitzer} images at spatial scales larger than 5 arc
seconds is evidence for a high redshift population. In Figure~\ref{fig-space}
we compare the spatial spectrum of the NICMOS residual fluctuations which we
have shown to be due to low redshift objects to the observed \emph{Spitzer}
fluctuations.  The fluctuations were normalized to be equal at 10 arc seconds.
Within the noise the spatial spectra are identical, removing the argument that
the observed spatial spectrum of the residual \emph{Spitzer} fluctuations 
require a high redshift origin.

\section{Conclusions}

We conclude that the Near Infrared Extragalactic Background has been resolved and
is due primarily to normal galaxies at redshifts near 1.  The claimed excess at
1.4 microns is false and arose from the inadequacy of zodiacal models to predict
the background level to the accuracy need to determine the true source flux.  We
further conclude that none of the properties of the fluctuation spectrum after
source subtraction in any of the 2MASS, NICMOS or \emph{Spitzer} images require
very high redshift ($z>10$) objects to account for them.

\acknowledgments

This article is based on data from observations with the NASA/ESA Hubble Space Telescope
obtained at the Space Telescope Science Institute, which is operated by the 
Association of Universities for Research in Astronomy under NASA contract
NAS 5-26555.

\end{document}